\documentclass[letterpaper, 10 pt, conference]{ieeeconf} 
\IEEEoverridecommandlockouts                              
\makeatletter

\let\proof\@undefined
\let\endproof\@undefined
\makeatother

\usepackage{amsmath} 
\usepackage{amssymb} 
\usepackage{amsthm}
\usepackage{dsfont}
\usepackage{mathrsfs}

\newcommand{\vect}[1]{\mathbf{#1}}
\newcommand{\mat}[1]{\mathbf{#1}}

\newcommand*\widebar[1]{\hbox{
    \vbox{%
      \hrule height 0.5pt 
      \kern0.5ex
      \hbox{%
        \kern-0.1em
        \ensuremath{#1}%
        \kern-0.1em
      }%
    }%
  }%
}

\newcommand{\colvec}[2][.9]{%
  \scalebox{#1}{%
    \renewcommand{\arraystretch}{.9}%
    $\begin{bmatrix}#2\end{bmatrix}$%
  }
}
\let\labelindent\relax
\usepackage{enumitem}

\usepackage{color}
\usepackage{float}
\usepackage{subfigure}
\usepackage{graphicx}
\usepackage{caption}
\DeclareMathAlphabet\mathbfcal{OMS}{cmsy}{b}{n}

\usepackage[usenames,dvipsnames,table]{xcolor}

\usepackage{hyperref} 
\hypersetup{
    colorlinks,
    citecolor=black,
    filecolor=black,
    linkcolor=black,
    urlcolor=black,
    pdfauthor={},
    pdfsubject={},
    pdftitle={}
}
\usepackage{multicol}
\usepackage{multirow}

\usepackage[ruled]{algorithm2e}
\usepackage{xcolor}
\usepackage{overpic}

\usepackage{cite}

\title{\LARGE \bf
Distributed Dual Quaternion Based Localization \\ 
of Visual Sensor Networks
}

\author{Luca Varotto, Marco Fabris, Giulia Michieletto, Angelo Cenedese
\thanks{All the authors are with the Department of Information Engineering,
	    Universit\`{a} degli Studi di Padova, Italy,
          Corresponding author: L. Varotto {\tt\small luca.varotto.5@phd.unipd.it}.}%
\thanks{This work was supported by the University of Padova under grant BIRD168152.}
}

\begin{document}

\maketitle
\thispagestyle{empty}
\pagestyle{empty}

\begin{abstract}
In this paper we consider the localization problem for a visual sensor network.
Inspired by the alternate attitude and position distributed optimization framework discussed in~\cite{TronVidal}, we propose an estimation scheme that exploits the unit dual quaternion algebra to 
describe the sensors pose. 
This representation 
is beneficial in the formulation of the optimization scheme allowing to solve the localization problem without designing two interlaced position and orientation estimators, thus improving the estimation error distribution over the two pose components and the overall localization performance. 
Furthermore, the numerical experimentation asserts the robustness of the proposed algorithm w.r.t. the initial conditions.


\end{abstract}

\section{Introduction}

A Visual Sensor Network (VSN) is a multi-sensor system composed of a collection of spatially distributed  camera-devices capable of communicating over a wireless network~\cite{kyung2016theory}.
Thanks to the advances 
in high performance embedded microcontrollers, optimized computer vision techniques and reliable communication protocols, multi-camera networks have increasingly spread
out in the last twenty years becoming ubiquitous smart systems in industrial, civil and domestic context. By acquiring information-rich data, these architectures enable vision-based interpretative applications as intelligent IoT surveillance~\cite{pavithra2017smart},
smart living domotics~\cite{gruenwedel2012decentralized},  autonomous vehicles assisted
driving~\cite{janai2017computer}, industrial environments monitoring and control~\cite{di2017vision}, urban perimeter/area patrolling for events detection~\cite{bof2017asynchronous}, to
cite a few. In all the mentioned scenarios,
the efficiency of most of the tasks is strictly conditioned by the knowledge from each device of its pose  w.r.t. a global inertial reference frame.

The self-estimation of the position and the attitude of each visual sensor node in a multi-device network corresponds to the \textit{localization} task. This constitutes a classic problem in the literature
on autonomous multi-camera systems (see, e.g.,~\cite{bajramovic2010self} and the references therein), mainly motivated by the fact that manual ad-hoc localization is a tedious and error prone task, not suitable to handle wide and/or dynamic networks.
, and becomes unfeasible with mobile camera networks. 
Consequently, the self-estimation of the position and the attitude of each visual sensor node in a multi-device network constitutes a classic problem in the literature
devoted to the autonomous multi-camera systems (see, e.g.,~\cite{hemayed2003survey},\cite{bajramovic2010self} and the references therein).

\noindent\textbf{Related works -} Various approaches have
been developed to solve the distributed VSNs localization problem. Most of these can be categorized according to the dimension of the estimation domain (planar vs. tridimensional), the
adopted solution paradigm (centralized vs. distributed), the available measurements (distances, angles of arrival, bearings), some a priori information (e.g., pose of some special nodes, referred as \textit{beacons} or \textit{anchors}) and some assumptions on the environment (e.g., presence of moving objects or markers). \\
From a mathematical perspective, 
the automated localization task can be recast as an optimization problem over a proper
 manifold. 
Some interesting results 
In this direction, 
the approach described 
in~\cite{TronVidal} consists in the iterative minimization of a suitable cost function through a distributed consensus-based strategy. In~\cite{knuth2013maximum} the previous method  is extended by employing maximum-likelihood estimation techniques. In~\cite{Ma} a generalized Newton scheme is proposed for a non-linear refinement of cameras pose exploiting the intrinsic Riemannian structure of the Essential manifold. Similarly, in~\cite{projectiveNewton}  a projective Newton optimization is performed on the Special Euclidean manifold $\mathbb{SE}(3)$.

In particular, the solution in~\cite{TronVidal} (hereafter also referred as \textit{TV algorithm}) represents the starting point of this work. 
This relies on the assumption that
each device in the network retrieves noisy relative pose measurements w.r.t. every other cameras with overlapping field of view (e.g., by employing standard computer vision techniques based on two-view geometry). 
These available measurements are then exploited to derive the cameras pose through a distributed minimization on $\mathbb{SE}(3)$ resting upon the least-squares (LS) principle. In particular, since $\mathbb{SE}(3)\! = \!\mathbb{SO}(3)\! \times\! \mathbb{R}^3$, the~optimization problem is split into two consecutive interlaced steps: first, a Riemannian gradient descent  is performed on the Special Orthogonal manifold $\mathbb{SO}(3)$ to estimate the cameras attitude, then the achieved estimates are used to derive the cameras position through a gradient descent on the Euclidean space~$\mathbb{R}^3$. One of the main drawbacks of this approach is that the estimation error is not equally distributed among the two pose components: because of the sequential scheme, the  position estimates accumulate much more error than orientation ones. Furthermore, the goodness of the final estimates strictly depends on the initial guess.

\noindent\textbf{Contributions -} To overcome the aforementioned issues, we adopt the (unit) dual quaternion formalism to unitarily represent the cameras pose. Through this convention, the LS cost function accounting for the pose estimation error 
needs to be suitably redefined, thus allowing the design of a single estimator for both position and orientation. 
Through this approach, our solution guarantees a more balanced estimation error  distribution as regards the two pose components and turns out to be more robust \mbox{w.r.t. the initial conditions.}
We also test the proposed algorithm assuming time-varying measurements. 
This scenario is strongly supported by the fact that a real VSN might be required to collect measurements over time to handle dynamic situations: this is the case, for instance, of mobile multi-camera systems subject to topological changes or \mbox{characterized by unpredictable dynamics.}

\noindent\textbf{Paper structure -} Sec.~\ref{sec: mathPreliminaries} is devoted to the recall of some mathematical preliminaries. In Sec.~\ref{sec: statement}, the localization problem is formally stated, providing the distributed dual quaternion based solution whose numerical results are reported in Sec.~\ref{sec: simulations}. Finally, the main conclusions are drawn in Sec.~\ref{sec: conclusions}.

\section{Mathematical Preliminaries}\label{sec: mathPreliminaries}

This section summarizes some principal foundations about quaternion~\cite{quaternions} and dual quaternion~\cite{DualQuaternions} algebras to provide a mathematical background for the proposed 
estimation law. 

\subsection{Quaternion Algebra} 

A  quaternion $\mathbf{q}$ is 
 an extension of a complex number
in a higher dimensional space. This is defined as the sum of a real and a complex part, i.e., \vspace{-0.25cm}
\begin{equation}
\mathbf{q} = \underbrace{q_0}_{\text{real part}} + \underbrace{\hat{\imath}q_1 + \hat{\jmath}q_2 + \hat{k}q_3}_{\text{complex part}} \in \mathbb{H},
\end{equation}
where $q_0, q_1, q_2, q_3  \in \mathbb{R}$, $\hat{\imath}, \hat{\jmath}, \hat{k}$ are such that $\hat{\imath}^2=\hat{\jmath}^2=\hat{k}^2= \hat{\imath}\hat{\jmath}\hat{k}=-1$, and $\mathbb{H}$ denotes the quaternion space. 
A quaternion $\mathbf{q} \in \mathbb{H}$ is generally represented as a four-dimensional vector 
composed by the \textit{scalar part} $q_0 \in  \mathbb{R}$ and the \textit{vector part} $\bar{\mathbf{q}} = [q_1 \; q_2 \; q_3]^\top \in \mathbb{R}^3$, i.e., $\mathbf{q} = [q_0 \;\; \bar{\mathbf{q}}^\top]^\top$.

Given two generic quaternions $\mathbf{p}, \mathbf{q}$, the conjugation and composition operations in  $\mathbb{H}$ are respectively defined as
\begin{align}
\mathbf{q}^\star &= \colvec{ q_0 & -\bar{\mathbf{q}}^\top}^\top,\\
\mathbf{p} \circ \mathbf{q} &= \colvec{p_0 q_0 - \bar{\mathbf{p}}^\top\bar{\mathbf{q}} & (p_0\bar{\mathbf{q}}+q_0\bar{\mathbf{p}} +\bar{\mathbf{p}}\times\bar{\mathbf{q}})^\top}^\top. \label{eq:quat_composition}
\end{align}
Notably, the composition rule~\eqref{eq:quat_composition} can be expressed as a matrix-vector product. Indeed, it holds that
\begin{align}
\mathbf{p} \circ \mathbf{q} 
& = \colvec{  p_0 & -\bar{\mathbf{p}}^\top \\ \bar{\mathbf{p}} & p_0 \mathbf{I}_3+[ \bar{\mathbf{p}} ]_\times }\colvec{q_0 \\ \bar{\vect{q}}} = {\mathbf{M}(\mathbf{p})} \mathbf{q}  \\
& = \colvec{ q_0 & -\bar{\mathbf{q}}^\top \\ \bar{\mathbf{q}} & q_0 \mathbf{I}_3-[ \bar{\mathbf{q}} ]_\times }\colvec{p_0 \\ \bar{\vect{p}}} = {\mathbf{N}(\mathbf{q})}\mathbf{p},
\end{align}
where $\mathbf{M}(\cdot),\mathbf{N}(\cdot) \in \mathbb{R}^{4 \times 4}$, $\mathbf{I}_3 \in \mathbb{R}^{3 \times 3}$ is the three dimensional identity matrix and $[\cdot]_\times$ indicates the map from $\mathbb{R}^3$ to the Lie algebra $\mathfrak{so}(3)$ that associates any non-zero vector 
to its corresponding skew-symmetric matrix.
%
As regards the quaternion composition, we also point out the next results that will be useful in the following:
\begin{equation}
\label{eq:composition}
\mathbf{p}^{\star} \circ \mathbf{q} = \mathbf{M}(\mathbf{p}^{\star})\mathbf{q} = \widetilde{\mathbf{N}}(\mathbf{q})\mathbf{p},
\end{equation}
where
\begin{equation}
\widetilde{\mathbf{N}}(\mathbf{q})= 
\colvec{
q_0 & \bar{\mathbf{q}}^\top\\ 
\bar{\mathbf{q}} & -q_0 \mathbf{I}_3+[\bar{\mathbf{q}}]_\times
} \in \mathbb{R}^{4\times 4}.
\end{equation}

To conclude, a quaternion $\vect{q} \in \mathbb{H}$ is called \textit{unit} when \mbox{$\Vert \vect{q} \Vert^2 = q_0^2+\bar{\vect{q}}^\top\bar{\vect{q}}=1$}. In this case, it belongs to the unit hypersphere $\mathbb{H}_u$ embedded in $\mathbb{R}^4$ and can be used to describe a rotation in 3D space. Formally, given two coordinates systems $\mathscr{F}_{i}$ and $\mathscr{F}_{j}$ such that $\mathscr{F}_{i}$ is rotated by an angle $\theta \in (-\pi,\pi]$ around the unit vector  $\vect{u} \in \mathbb{R}^3$ w.r.t. $\mathscr{F}_{j}$, their relative orientation can be represented by the unit quaternion 
\begin{align}
\label{eq:quaternion_axis_angle}
\vect{q} = \colvec{ \cos \frac{\theta}{2} & \vect{u}^\top \sin \frac{\theta}{2}}^\top \in \mathbb{H}_u.
\end{align}


%
%

\subsection{Dual Quaternion Algebra}


Within the dual number theory context, a dual quaternion $\vect{d}$ is generally defined as
\begin{equation}
\mathbf{d} = \mathbf{q}_r + \epsilon \mathbf{q}_d \in \mathbb{DH},
\end{equation}
where the (single) quaternions $\mathbf{q}_r, \mathbf{q}_d \in \mathbb{H}$ represent its \textit{real} and \textit{dual parts}, $\epsilon$ is the nilpotent dual unit satisfying $\epsilon^2=0$ and $\epsilon \neq 0$ and $\mathbb{DH}$ denotes the dual quaternion space. 

Considering $\vect{d}=\mathbf{q}_r + \epsilon \mathbf{q}_d$ and $\vect{d}'=\mathbf{q}_r' + \epsilon \mathbf{q}_d'$ in $\mathbb{DH}$, the composition operation on this manifold is given by
\begin{align}
\vect{d}\odot \vect{d}'&\!=\!\vect{q}_r \circ \vect{q}_r' \!+\!\epsilon (\vect{q}_r \circ \vect{q}_d' + \vect{q}_d \circ \vect{q}_r'). \label{eq:dual_quat_composition}
\end{align}

From a mathematical perspective, a dual quaternion can be suitably represented as an eight-dimensional vector, e.g., $\vect{d} = [q_{r,0} \; \bar{\vect{q}}_r^\top \; q_{d,0} \; \bar{\vect{q}}_d^\top ]^\top$, where $q_{r,0}, q_{d,0} \in \mathbb{R}$ 
and $\bar{\vect{q}}_r, \bar{\vect{q}}_d \in \mathbb{R}^3$ according to the single quaternion notation. Hence, similarly to the quaternion case, the composition~\eqref{eq:dual_quat_composition} can be expressed as matrix-vector multiplication, namely
\begin{equation}
\begin{split}
\mathbf{d} \odot \mathbf{d}^\prime 
& = \colvec{ \mathbf{M}(\mathbf{q}_r) & \mathbf{0}_{4} \\ \mathbf{M}(\mathbf{q}_d) & \mathbf{M}(\mathbf{q}_r)} \colvec{\vect{q}_r' \\  \vect{q}_d'} = \mathbf{U}(\mathbf{d}) \mathbf{d}^\prime\\
&=  \colvec{ \mathbf{N}(\mathbf{q}_r^\prime) & \mathbf{0}_{4} \\ \mathbf{N}(\mathbf{q}_d^\prime) & \mathbf{N}(\mathbf{q}_r^\prime) }  \colvec{\vect{q}_r \\  \vect{q}_d} ={\mathbf{V}(\mathbf{d}^\prime)} \mathbf{d}.
\end{split}
\end{equation}
where $\mat{U}(\cdot), \mat{V}(\cdot) \in \mathbb{R}^{8 \times 8}$ and the notation $\mathbf{0}_{4}$ is used to indicate a ($4 \times 4$) matrix whose entries are  all zeros.
Note that, analogously to~\eqref{eq:composition}, it also holds that
\begin{equation}
\mathbf{d}^{\star} \odot \mathbf{d}^\prime = \mathbf{U}(\mathbf{d}^{\star})\mathbf{d}^\prime = \widetilde{\mathbf{V}}(\mathbf{d}^\prime)\mathbf{d},
\end{equation}
where $\mathbf{d}^\star = \mathbf{q}_r^\star + \epsilon \mathbf{q}_d^\star \in \mathbb{DH}$ is the conjugate of the dual quaternion $\vect{d} \in \mathbb{DH} $ and $\widetilde{\mathbf{V}}(\mathbf{d}) \in \mathbb{R}^{8 \times 8}$ is defined as 
\begin{equation}
\widetilde{\mathbf{V}}(\mathbf{d})= 
\colvec{
\widetilde{\mathbf{N}}(\mathbf{q}_r) & \mathbf{0}_{4 } \\ 
\widetilde{\mathbf{N}}(\mathbf{q}_d) & \widetilde{\mathbf{N}}(\mathbf{q}_r)
}.
\end{equation}

Finally, a dual quaternion  $\mathbf{d} \in \mathbb{DH}$ is called \textit{unit} when \mbox{$\lVert \mathbf{q}_r \rVert^2\!=\!1$} (its real part is a unit quaternion) and \mbox{$\mathbf{q}_r^\top \mathbf{q}_d = 0$} (its real and dual part are orthogonal).
In this case, the dual quaternion belongs to the set $\mathbb{DH}_u$ and can be used to describe a pose transformation in 3D space. Formally, given two coordinates systems $\mathscr{F}_i$ and $\mathscr{F}_j$,  their relative orientation $\mathbf{q}_r \in \mathbb{H}_u$ and the position $\mathbf{p} \in \mathbb{R}^3$ of the origin of $\mathscr{F}_i$ in $\mathscr{F}_j$ (given in $\mathscr{F}_j$) can be represented by the dual quaternion 
\vspace{-0.25cm}
\begin{equation}\label{eq:RTtoDH}
\mathbf{d} = \mathbf{q}_r + \frac{\epsilon}{2} \, \mathbf{q}_t \circ \mathbf{q}_r \in \mathbb{DH}_u,\vspace{-0.25cm}
\end{equation} 
where $\mathbf{q}_t = \colvec{0 & \mathbf{p}^\top}^\top  \in \mathbb{H}$ is the (pure) quaternion embedding the vector  $\mathbf{p}\in \mathbb{R}^3$.


\section{Distributed Localization of a VSN}\label{sec: statement}

In this section the localization problem for a VSN is formally stated and, inspired by~\cite{TronVidal}, a suitable distributed solution is provided resting upon a gradient descent minimization. The original and innovative aspect relies on the cameras pose representation via the dual quaternion formalism and its exploitation in the optimization framework for the design of the localization algorithm.   

\subsection{Problem Statement}

Consider a VSN made up of $n$ cameras. According to the most agreed pinhole model~\cite{Ma}, we assume that each device $i \in \{ 1 \ldots n\}$  is associated to a local frame $\mathscr{F}_i = \{O_i,(X_i,Y_i,Z_i)\}$ so that the origin $O_i$ coincides with the center of
projection of the camera, while the $Z_i$-axis (named \textit{optical axis}) is pointed in
the viewing direction, the $Y_i$-axis is up faced and the $X_i$-axis is oriented according to the left-handed coordinates system. The pose $g_i \in \mathbb{SE}(3)$ of the camera is thus identified by the position and the orientation of $\mathscr{F}_i$ w.r.t. the global inertial frame. 

We then model the network  as a connected undirected graph $\mathcal{G} = (\mathcal{V},\mathcal{E})$ so that the $i$-th device is identified by $i$-th node in the vertex set $\mathcal{V}$ and the edge set \mbox{$\mathcal{E}\subseteq\mathcal{V}\times \mathcal{V}$} is determined by accounting for the network visibility constraints, meaning that $(i,j) \in \mathcal{E}$ if the $i$-th and $j$-th cameras have (partially) overlapping field of view. For the sake of simplicity, we also assume that the neighboring cameras (corresponding to adjacent nodes in the graph) are  able to communicate according to a preset protocol. This means that $\mathcal{E}$ describes both the cameras \mbox{communication and sensing interactions.} 

The distributed localization problem consists in the determination of the pose (w.r.t. the inertial frame) of each  camera composing the given VSN, by exploiting the available measurements, in a distributed manner.   
To this end, it is useful to observe that for each pair of cameras $i$ and $j$ with overlapping fields of view (i.e., such that $(i,j) \in \mathcal{E}$),
the pose $g_i$ of the $i$-th camera is linked to the pose $g_j$ of the neighboring $j$-th
camera. In fact, the relative change of coordinates
$g_{ij} = g_i^{-1} \circ g_j$ holds
and, consistently, the pose of camera $j$ w.r.t. that of $i$ results as $g_j = g_i \circ g_{ij}$. 
In the following, $g_{ij}$ is referred as the \textit{relative} pose between the $i$-th and the $j$-th neighboring cameras, whereas the pose $g_i$ of camera $i$ is said \textit{absolute}, and represents its state.
Moreover, each camera in the network is assumed to be able to employ standard computer vision algorithms (see, e.g.,~\cite{8points},~\cite{DLT},~\cite{Zhang}) to recover a noisy measurement  $\widetilde{g}_{ij} \in \mathbb{SE}(3)$ of the relative pose w.r.t. to any device in its neighborhood $\mathcal{N}_i = \{ j \in \mathcal{V} \; \vert \; (i,j) \in \mathcal{E}\}$. Note that in general, it occurs that $\widetilde{g}_{ij} \neq \left(\widetilde{g}_{ji}\right)^{-1}$, due to noise effects.

In the light of these observations, the main objective of the localization task for a given VSN represented by $\mathcal{G} = (\mathcal{V}, \mathcal{E})$  is the computation of a set of the relative transformations $\{ \hat{g}_{ij}, (i,j) \in \mathcal{E} \}$ that satisfy the consistency constraint along network cycles\footnote{The consistency constraint requires the transformation along any cycle in the network results to be identical.}
 and, simultaneously, are as close as possible to the available measurements $\{ \widetilde{g}_{ij}, (i,j) \in \mathcal{E} \}$. To this aim,  employing a LS approach, it is convenient to minimize the following cost function
\begin{equation}
\rho(\{\hat{g}_{ij}\}) = \textstyle\sum\limits_{i \in \mathcal{V}} \sum\limits_{j \in \mathcal{N}_i} \frac{1}{2} d_g^2(\hat{g}_{ij},\widetilde{g}_{ij}),
\end{equation}
where $d_g(\cdot,\cdot)$ defines the distance on the $\mathbb{SE}(3)$ manifold. Note that the available measurements may not be consistent.
To enforce this constraint, the relative transformations can be reparametrized using the absolute ones, i.e.,
\begin{equation}
\label{eq:cost_fun_0}
\rho(\{ \hat{g}_i \}) = \textstyle\sum\limits_{i \in \mathcal{V}} \sum\limits_{j \in \mathcal{N}_i} \frac{1}{2} d_g^2(\hat{g}_i^{-1} \circ \hat{g}_j,\widetilde{g}_{ij}).
\end{equation}
In particular, 
we observe that the cost function~\eqref{eq:cost_fun_0} can be rewritten in a form more convenient for the application of a distributed paradigm, namely
\begin{align}
\label{eq:cost_fun}
\rho(\{ \hat{g}_i \})&=\textstyle\sum\limits_{i \in \mathcal{V}} \sum\limits_{j \in \mathcal{N}_i}  \frac{1}{4}  \left( d_g^2(\hat{g}_i^{-1} \circ \hat{g}_j,\widetilde{g}_{ij}) +  d_g^2(\hat{g}_j^{-1} \circ \hat{g}_i,\widetilde{g}_{ji})    \right)\nonumber \\
&=\textstyle\sum\limits_{i \in \mathcal{V}} \rho_i(\hat{g}_i). 
\end{align}

It is straightforward that the self-localization problem can be solved in a distributed way: each camera is~required to solve the (non-linear) minimization of the
local cost function $\rho_i(\hat{g}_i)$. This depends only on information  available locally or through 1-hop communication: the actual pose estimates $\hat{g}_i$ and $\{ \hat{g}_{j}$, $j \in \mathcal{N}_i \}$, and the measurements~sets $\{\tilde{g}_{ij}, j \in \mathcal{N}_i\}$ and  $\{\tilde{g}_{ji}, j \in \mathcal{N}_i\}$. In particular, this optimization can be achieved using an iterative consensus framework, as in~\cite{TronVidal}.

\subsection{Dual Quaternion Based Solution}

In~\cite{TronVidal}, the minimization of the cost function~\eqref{eq:cost_fun} is addressed by exploiting the fact that $\mathbb{SE}(3) = \mathbb{SO}(3) \times \mathbb{R}^3$. 
Each pose $g_i \in \mathbb{SE}(3)$ corresponds to the pair $(\mat{R}_i, \vect{p}_i)$ where $\mat{R}_i \in \mathbb{SO}(3)$ is a rotation matrix and $\vect{p}_i \in \mathbb{R}^3$ is a position vector. The distance function $d_g(\cdot,\cdot)$ can then be redefined by considering the metrics on $\mathbb{SO}(3)$ and $\mathbb{R}^3$ as 
\begin{align}
\label{eq:metrics_SE2}
d_g^2(\hat{g}_{ij},\widetilde{g}_{ij}) &=  d_{\mathbb{SO}(3)}^2(\hat{\mathbf{R}}_{ij},\widetilde{\mathbf{R}}_{ij}) \!+\!  \lVert \hat{\mathbf{R}}_i^\top\hat{\mathbf{p}}_{ij}-\widetilde{\mathbf{p}}_{ij} \rVert^2 \\
& = d_{\mathbb{SO}(3)}^2(\hat{\mathbf{R}}_i^\top \hat{\mathbf{R}}_j,\widetilde{\mathbf{R}}_{ij}) \!+ \! \lVert \hat{\mathbf{R}}_i^\top(\hat{\mathbf{p}}_j-\hat{\mathbf{p}}_i)-\widetilde{\mathbf{p}}_{ij} \rVert^2 \nonumber
\end{align}
and consequently, substituting~\eqref{eq:metrics_SE2} in~\eqref{eq:cost_fun}, it follows that
\begin{align}
\label{eq:cost_TV}
\rho(\{ \hat{g}_i \}) = {\rho_R(\{\hat{\mathbf{R}}_i\})} + {\rho_T(\{\hat{\mathbf{R}}_i,\hat{\mathbf{p}}_i\})}.
\end{align}

The TV algorithm described in~\cite{TronVidal} 
solves the distributed localization task by first minimizing $\rho_R(\cdot)$ through a Riemannian gradient descent and then $\rho_T(\cdot)$ applying a Euclidean gradient descent, thus splitting the optimization problem in two consequent steps. Following this scheme, it is perceivable that the estimation error accumulated during the first step (minimization of $\rho_R(\cdot)$) is inherited by the second step (minimization of $\rho_T(\cdot)$), hence, the estimation inaccuracy is not equally distributed over the two pose components. Moreover, it is verified that the performance of the optimization procedure strongly depends on the estimates initialization. 
 
 To overcome all the aforementioned issues, we intend to  reformulate the cost function~\eqref{eq:cost_fun} using the algebra of unit dual quaternions, although this implies to consider $\mathbb{DH}_u$ instead of $\mathbb{SE}(3)$ as poses domain. To this end, we consider the following distance function
\begin{align}
d_{\mathbb{DH}_u}(\hat{\mathbf{d}}_{ij},\widetilde{\mathbf{d}}_{ij}) &= \Vert \widetilde{\mathbf{d}}_{ij}^{\star} \odot \hat{\mathbf{d}}_{ij} \Vert = \Vert \widetilde{\mathbf{d}}_{ij}^{\star} \odot (\hat{\mathbf{d}}_i^{\star} \odot \hat{\mathbf{d}}_j)\Vert \\
& = \Vert \mathbf{U}(\widetilde{\mathbf{d}}_{ij}^{\star})\widetilde{\mathbf{V}}(\hat{\mathbf{d}}_j)\hat{\mathbf{d}}_i\Vert, \label{eq:KCSmetric}
\end{align}
where $\hat{\mathbf{d}}_{ij},\widetilde{\mathbf{d}}_{ij} \in \mathbb{DH}_u$  denote the estimate and the measurement of the relative pose between the $i$-th and the \mbox{$j$-th} neighboring cameras, respectively, while $\hat{\mathbf{d}}_i, \hat{\mathbf{d}}_j \in \mathbb{DH}_u$~indicate the estimate of their absolute poses.

Employing~\eqref{eq:KCSmetric}, the cost function~\eqref{eq:cost_fun} turns out to be
\begin{align}
\label{eq:KCSfunctional}
&\rho(\{\hat{\mathbf{d}}_i\})
=\textstyle\sum\limits_{i \in \mathcal{V}} \rho_i(\hat{\mathbf{d}}_i), \qquad \text{where}\\
%
&\rho_i(\hat{\mathbf{d}}_i)  \!=\! \! \textstyle\sum\limits_{j \in \mathcal{N}_i} \!\!  \frac{1}{4} \!\! \left(\lVert  \mathbf{U}(\widetilde{\mathbf{d}}_{ij}^{\star})\widetilde{\mathbf{V}}(\hat{\mathbf{d}}_j)\hat{\mathbf{d}}_i  \rVert ^2 \!+\!  \lVert  \mathbf{U}(\widetilde{\mathbf{d}}_{ji}^{\star})\widetilde{\mathbf{V}}(\hat{\mathbf{d}}_i)\hat{\mathbf{d}}_j  \rVert ^2 \right).\nonumber 
\end{align}
%
%

\begin{algorithm}[t!]
{\small
 $\quad$\\
  \tcp{$t_{max}$ total number of iterations}
 \tcp{$n$ number of cameras in the network}
 
 \textbf{Input}: relative noisy measurements $\{\widetilde{\mathbf{d}}_{ij}, (i,j) \in \mathcal{E}\}$\\
 
  \textbf{Initialization} ($t=0$): 
  
  initial pose estimates $\{\hat{\mathbf{d}}_i(0), i \in \mathcal{V} \}$, $\hat{\mathbf{d}}_1(0)= \mathbf{d}_I$ \\ initial cost function value $\rho(0)= \rho (\{\hat{\mathbf{d}}_i(0)\},\{\widetilde{\mathbf{d}}_{ij}\})$ \\
  
  \For{$t=1:t_{max}$}{$\widehat{\mathbf{d}}_1(t)= \mathbf{d}_I$
  
  	  \For{$i=2:n$}{

  	   		\textbf{Update}: $\hat{\mathbf{d}}_i(t) = \hat{\mathbf{d}}_i(t-1) - \delta \frac{\partial \rho_i(t-1)}{\partial \hat{\mathbf{d}}_i(t-1)}, \quad \delta > 0$\\
  	   		
  	   		\textbf{Normalization}: $\hat{\mathbf{d}}_i(t) = \hat{\mathbf{q}}_{r,i}(t)+ \epsilon \hat{\mathbf{q}}_{d,i}(t)$\\$\rightarrow \hat{\mathbf{d}}_i'(t) = \hat{\mathbf{q}}_{r,i}'(t)+ \epsilon \hat{\mathbf{q}}_{d,i}'(t)$ \quad with \\
  	   		\hspace{0.5cm}
$
\hat{\vect{q}}_{r,i}'(t)= \frac{\hat{\vect{q}}_{r,i}(t)}{\lVert \hat{\vect{q}}_{r,i}(t) \rVert }$ \\

\hspace{0.5cm}
$  		 \hat{\vect{q}}_{d,i}'(t) = \hat{\vect{q}}_{d,i}(t)  - \hat{\vect{q}}_{r,i}(t)(\hat{\vect{q}}_{d,i}'(t)^\top \hat{\vect{q}}_{r,i}(t))
$  	   		 
       }
     $\rho(t)= \rho (\{\hat{\mathbf{d}}_i'(t)\}, \{\widetilde{\mathbf{d}}_{ij}\})$  
  }
}
 \caption{DDQL}
 \label{algo:DQreformulation}
\end{algorithm}

Adopting the dual quaternion formalism,  the minimization of~\eqref{eq:KCSfunctional} can still be performed following the distributed paradigm; in particular, each device in the network can locally exploit 
an iterative  optimization strategy to minimize $\rho_i(\hat{\mathbf{d}}_i)$. Hence, we suppose that, after an initialization step required to set the initial pose estimates, at every iteration $t$ each node $i$ in the network performs the following procedure. 
\begin{enumerate}[leftmargin=0.5cm]
\item Denoting the current estimate of the absolute pose $\vect{d}_i$ by $\hat{\vect{d}}_i(t)$, it first determines the derivative of $\rho_i(t) =\rho_i(\hat{\mathbf{d}}_i(t))$ w.r.t. $\hat{\mathbf{d}}_i(t)$, whose expression is given by
\begin{equation*}
\begin{split}
\dfrac{\partial \rho_i(t)}{\partial  \hat{\mathbf{d}}_i(t)}  
& \!= \! \textstyle\sum\limits_{j \in \mathcal{N}_i} \left( \widetilde{\mathbf{V}}(\hat{\mathbf{d}}_j(t))^\top \mathbf{U}(\widetilde{\mathbf{d}}_{ij}^{\star})^\top \mathbf{U}(\widetilde{\mathbf{d}}_{ij}^{\star}) \widetilde{\mathbf{V}}(\hat{\mathbf{d}}_j(t))+ \right. \\
&  \;\;\; \left. + \mathbf{U}(\hat{\mathbf{d}}_j ^{\star}(t))^{\!\top} \mathbf{U}(\widetilde{\mathbf{d}}_{ji}^{\star})^{\!\top} \mathbf{U}(\widetilde{\mathbf{d}}_{ji}^{\star}) \mathbf{U}(\hat{\mathbf{d}}_j ^{\star}(t)) \right)\! \hat{\mathbf{d}}_i(t).
\end{split} 
\end{equation*}
\item Then, it determines the estimation $\hat{\vect{d}}_i (t + 1)$ related to the subsequent iteration by performing a steepest gradient descent step, so that
\begin{align}
\label{eq:update}
\hat{\vect{d}}_i (t + 1) = \hat{\vect{d}}_i (t) - \delta  \dfrac{\partial \rho_i(t)}{\partial  \hat{\mathbf{d}}_i(t)},
\end{align}
where $\delta >0 $ constitutes a properly chosen step-size.
\item  Since the additive update rule~\eqref{eq:update} may violate the constraints of $\mathbb{DH}_u$ space, a normalization step is performed on the computed \mbox{$\hat{\vect{d}}_i(t + 1)=\hat{\vect{q}}_{r,i}(t + 1)+\epsilon \hat{\vect{q}}_{d,i}(t + 1)$}. The new~estimates thus results $\hat{\vect{d}}_i'(t + 1)=\hat{\vect{q}}_{r,i}' (t + 1)+\epsilon \hat{\vect{q}}_{d,i}' (t + 1)$
with
\begin{align}
&\hat{\vect{q}}_{r,i}'= \dfrac{\hat{\vect{q}}_{r,i}}{\lVert \hat{\vect{q}}_{r,i} \rVert }, 
 \quad \hat{\vect{q}}_{d,i}' = \hat{\vect{q}}_{d,i}  - \hat{\vect{q}}_{r,i}(\hat{\vect{q}}_{d,i}'^\top \hat{\vect{q}}_{r,i}),
\end{align}
dropping the time dependency for the sake of compactness.
The pose estimate $\hat{\vect{d}}_i'(t + 1)$ is finally communicated to all the neighboring nodes $j \in \mathcal{N}_i$.

\end{enumerate}

The iterative algorithm stops after a preset number $t_{max} \in \mathbb{N}$ of iterations, which
has to be large enough to guarantee the achievement of a minimum for the cost function.

Alg.~\ref{algo:DQreformulation} summarizes the described procedure (referred as \textit{Distributed Dual Quaternion based Localization}, DDQL) that allows to estimate the absolute pose of each camera in a VSN starting from the noisy and inconsistent pose measurements. Note that, w.l.o.g., we assume that these are expressed according to the dual quaternion formalism.  Moreover, 
we suppose that the body frame of camera 1 coincides with the global reference frame, namely its pose is always encoded by $\mathbf{d}_I \in \mathbb{DH}_u$, i.e., the dual quaternion corresponding to identical rotation and the zero position. This assumption is necessary to solve the localization problem when no a priori knowledge on the pose of a camera is available and it implies that all the final estimates are biased by the same
quantity w.r.t. to the inertial frame.


\section{Multi-sampling Measurements}\label{sec: multiMeas}

In the previous section, we have supposed that the measurements are invariant over time, i.e., they are collected only once and used in a static scenario for the localization optimization problem. In this section, accounting for a dynamic scenario involving VSNs subject to configuration changes,  we assume to deal with time-varying measurements. In particular, we do not account for topological variation (i.e., the network graph is invariant) but we suppose that the relative pose measurements continuously change according to a certain rate, namely after \mbox{$0<T_s<t_{max}$} iterations. 

In detail, we assume that a set of relative measurements is available at $t=0$ and a first localization step is performed, then at each iteration $t = kT_s$, $k\geq1$  new measurements are supposed to be collected so that a new localization step begins. The resulting estimates of  the previous step (corresponding to the estimates at the iteration $t=kT_s-1$) are exploited in the initialization. During the interval $[kT_s, (k+1)T_s-1]$, the pose estimates are then updated accordingly to the strategy summarized in Alg.~\ref{algo:DQreformulation}. The solution of the distribution localization problem thus corresponds to the final estimates of the last step of the described procedure that is recapped in Alg.~\ref{algo:multiMeas}.

\begin{algorithm}[t]
{\small
\tcp{$t_{max}$ total number of iterations}

 \textbf{Input}: relative noisy measurements $\{\widetilde{\mathbf{d}}_{ij}(0), (i,j) \in \mathcal{E}\}$\\
 
  \textbf{Initialization} ($t=0$): 
  
  initial pose estimates $\{\widehat{\mathbf{d}}_i(0), i \in \mathcal{V} \}$, $\widehat{\mathbf{d}}_1(0)= \mathbf{d}_I$  \\ initial cost function value $\rho(0)= \rho (\{\widehat{\mathbf{d}}_i(0)\},\{\widetilde{\mathbf{d}}_{ij}(0)\})$ \\ 
  initial counter value $k=0$\\

 \For{$t=1:t_{max}$}{ 
 	\If{$t\mathrm{mod}T_s== 0$}{
 	 	update counter $k \leftarrow k+1$\\
 		new measurements $\{\widetilde{\mathbf{d}}_{ij}(k T_s), (i,j) \in \mathcal{E}\}$
 	}
 
		Alg.~\ref{algo:DQreformulation} with \\ 
		 input $\{\widetilde{\mathbf{d}}_{ij}(kT_s), (i,j) \in \mathcal{E}\}$\\
		   initial estimates {\footnotesize $\{\widehat{\mathbf{d}}_i(max\{0,kT_s-1\}), i \in \mathcal{V} \}$} \\
		   initial cost function value {\footnotesize $\rho (\{\widehat{\mathbf{d}}_i(max\{0,kT_s-1\})\},\{\widetilde{\mathbf{d}}_{ij}(kT_s)\})$}\\

 }
 }
\caption{DDQL: multi-sampling measurements}
 \label{algo:multiMeas}
\end{algorithm}

To discuss the performance of the algorithm two  parameters have to be considered: the value of $T_s$ and the variance of the measurements noise. If $T_s$ is too small, it may happen that the localization steps (corresponding to the implementation of Alg.~\ref{algo:DQreformulation}) are not long enough to sensitively improve the estimates, causing a very slow convergence. If $T_s$ is too large, instead, during each step the estimates have enough time to converge to the actual target and, when new measurements are acquired, the new target might cause an oscillatory trend of the cost function, therefore  the system may become unstable. A good choice of the $T_s$ parameter is thus necessary. In addition, if the relative measurements are very noisy, there exist chances for the target to be reached; however, the pose estimates computed at $t=kT_s-1$ could not be good initial conditions for the successive localization step, impairing the cost function convergence. 

%
%
%


\section{Numerical Results}\label{sec: simulations}

In this section some simulative results are provided to show the performance of the  DDQL algorithm. Its effectiveness is confirmed by the comparison with the TV algorithm described in~\cite{TronVidal}, focusing on the advantages carried out by the cost function reformulation  through the dual quaternion formalism and the optimal solution that follows.

\subsection{Performance of DDQL Algorithm}
\begin{figure}[!t]
\centering
    \subfigure[]{
    \includegraphics[width=0.36\textwidth]{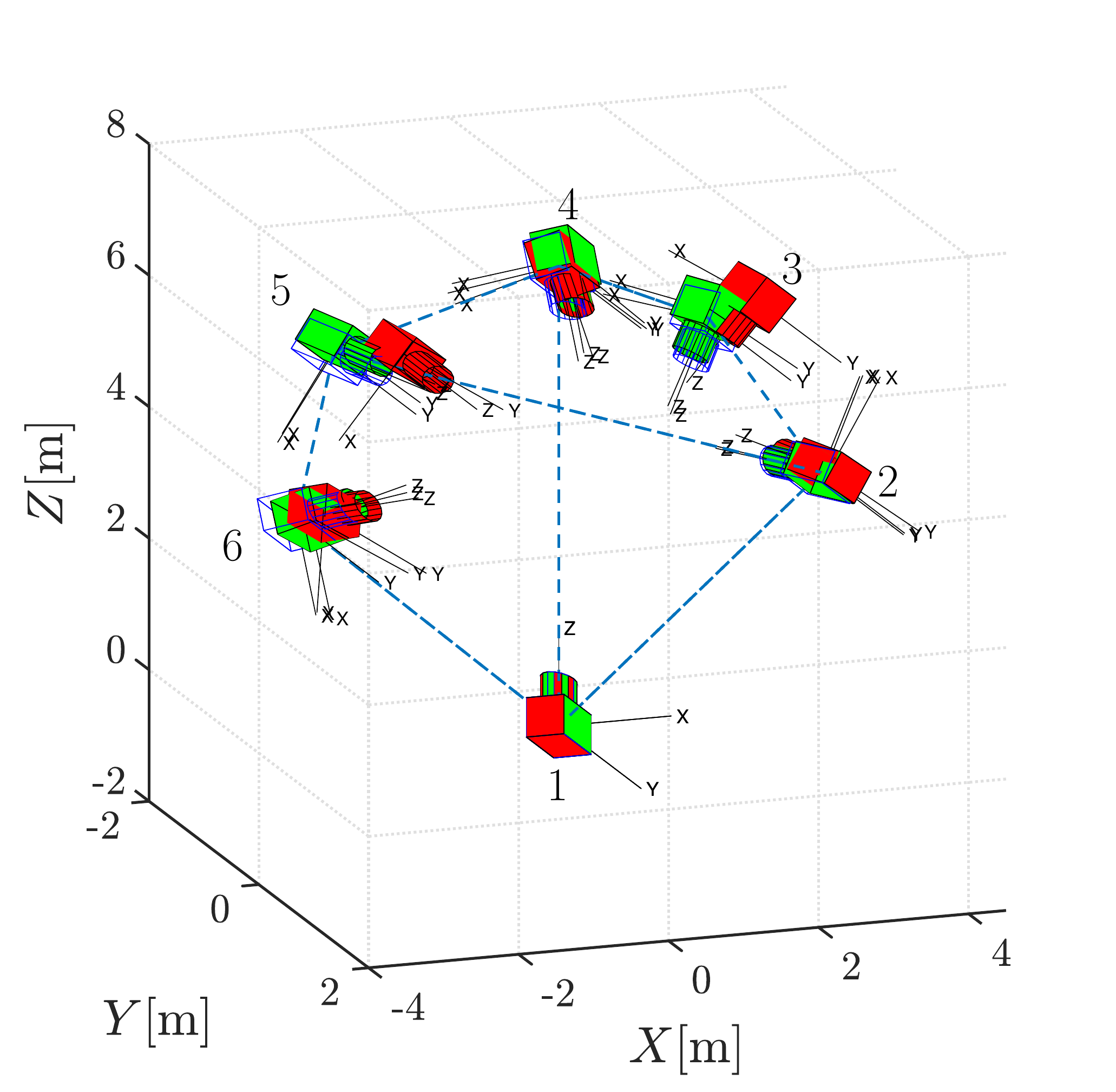} \label{fig:standardCase_a}
}\vspace{-0.38cm}
    \subfigure[]{
   \includegraphics[scale=0.25]{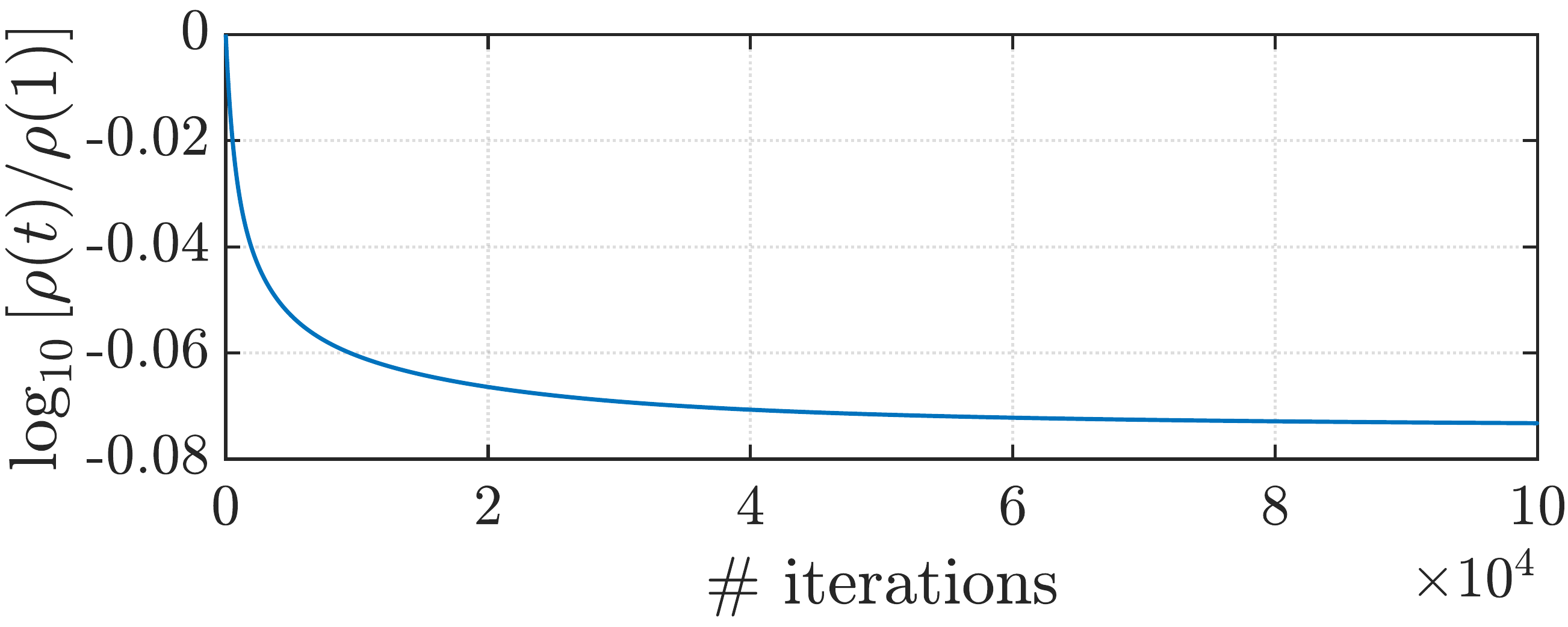} 
   \label{fig:standardCase_b}
   }  
\caption{Performance of DDQL algorithm: (a) real cameras poses (blue), initial estimates (red) and final estimates (green) of the Alg.~\ref{algo:DQreformulation}; (b) cost function trend in logarithmic scale.}
\vspace{-0.5cm}
\label{fig:standardCase}
\end{figure}

To illustrate the effectiveness of Alg.~\ref{algo:DQreformulation}, we consider  the camera network in Fig.~\ref{fig:standardCase_a}  composed of $n=6$ devices located at the same height from the ground, i.e., a planar VSN that can be used, for instance, for monitoring a certain area. The simulation campaign is carried out accounting for the cameras interactions defined by the graph depicted in light blue (dashed lines) in the aforementioned figure, where we also report the real absolute cameras poses, namely the  
 ground truth of the test. 
 We generate the noisy relative measurements by composing the real absolute poses corrupted by noise. In detail, the (absolute) orientations are altered by additional white Gaussian noise with standard deviation $\sqrt{5}$deg on the $X$ and $Z$-axis (tilt and roll angles) and $5$deg on the $Y$-axis (pan angle), whereas a white Gaussian noise with standard deviation $\sqrt{0.005}$m along all the axes is added to the (absolute) positions.   
The DDQL algorithm is run by setting $\delta =  10^{-4}$ and $t_{max}=10^5$ iterations, and assuming to start by the initial pose estimates marked in red in Fig.~\ref{fig:standardCase_a}. 

\begin{figure}[!t]
\begin{center}
\subfigure[]{
    \includegraphics[width=0.36\textwidth]{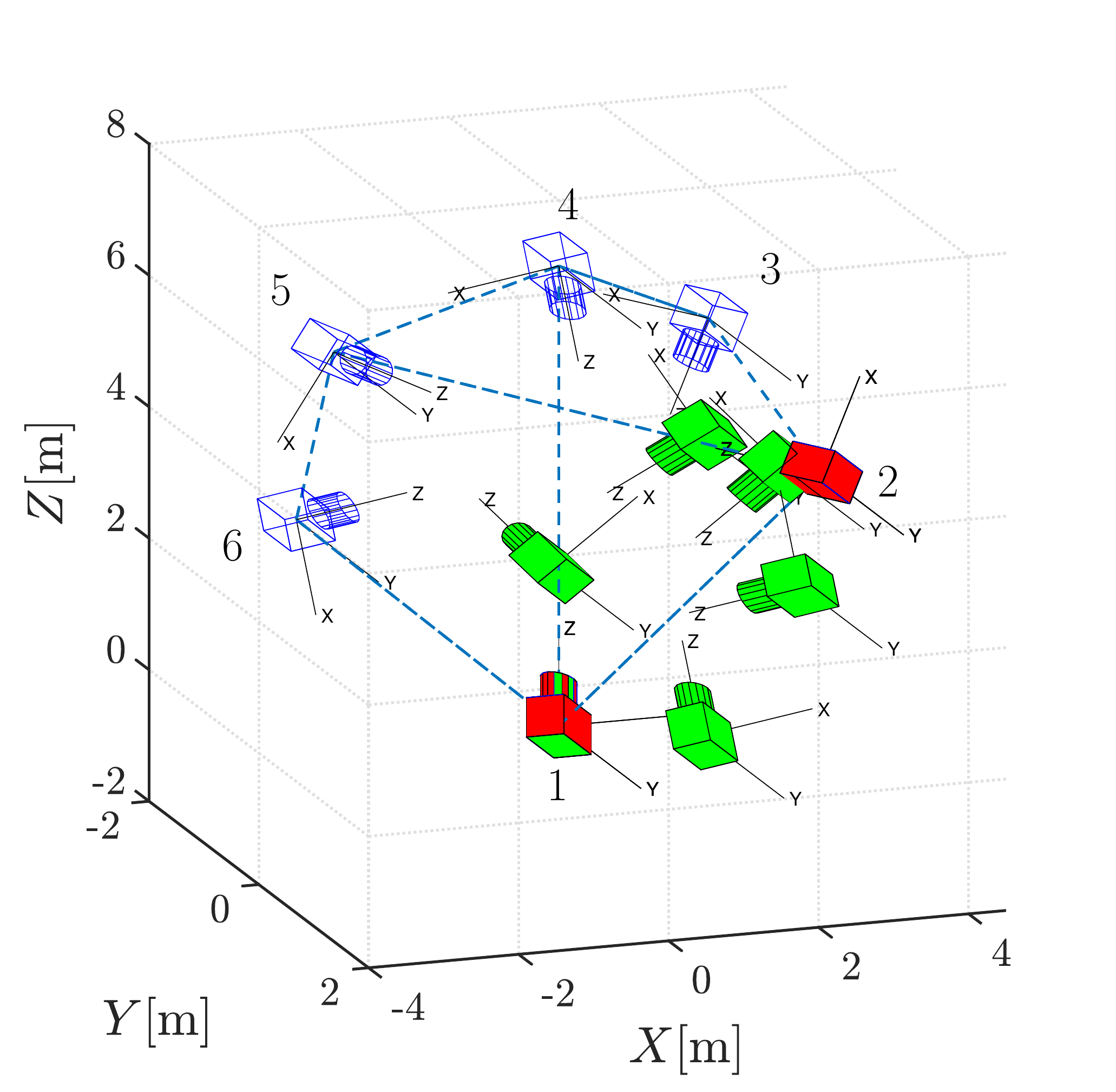}\label{fig:TVBadCI_a}}\vspace{-0.36cm}
\subfigure[]{
\includegraphics[scale=0.3]{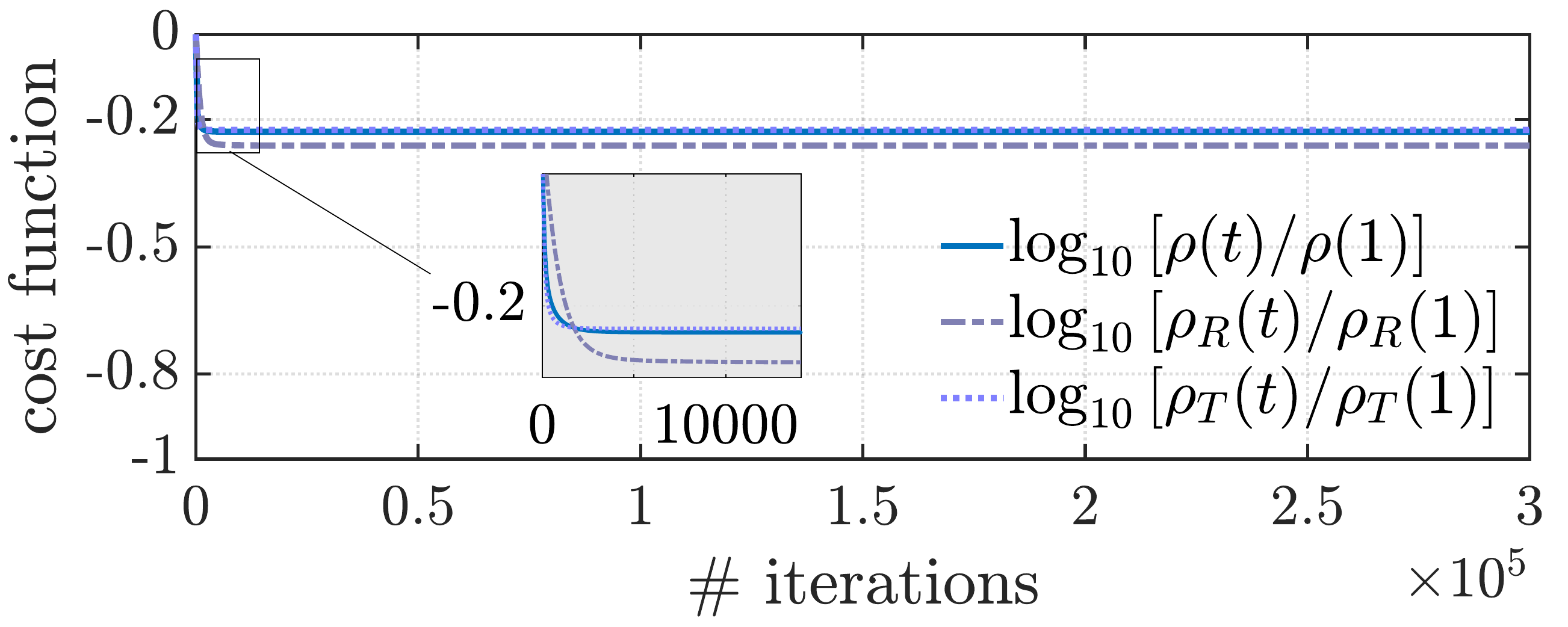}
\label{fig:TVBadCI_b}}\vspace{-0.36cm}
\subfigure[]{
\includegraphics[scale =0.3]{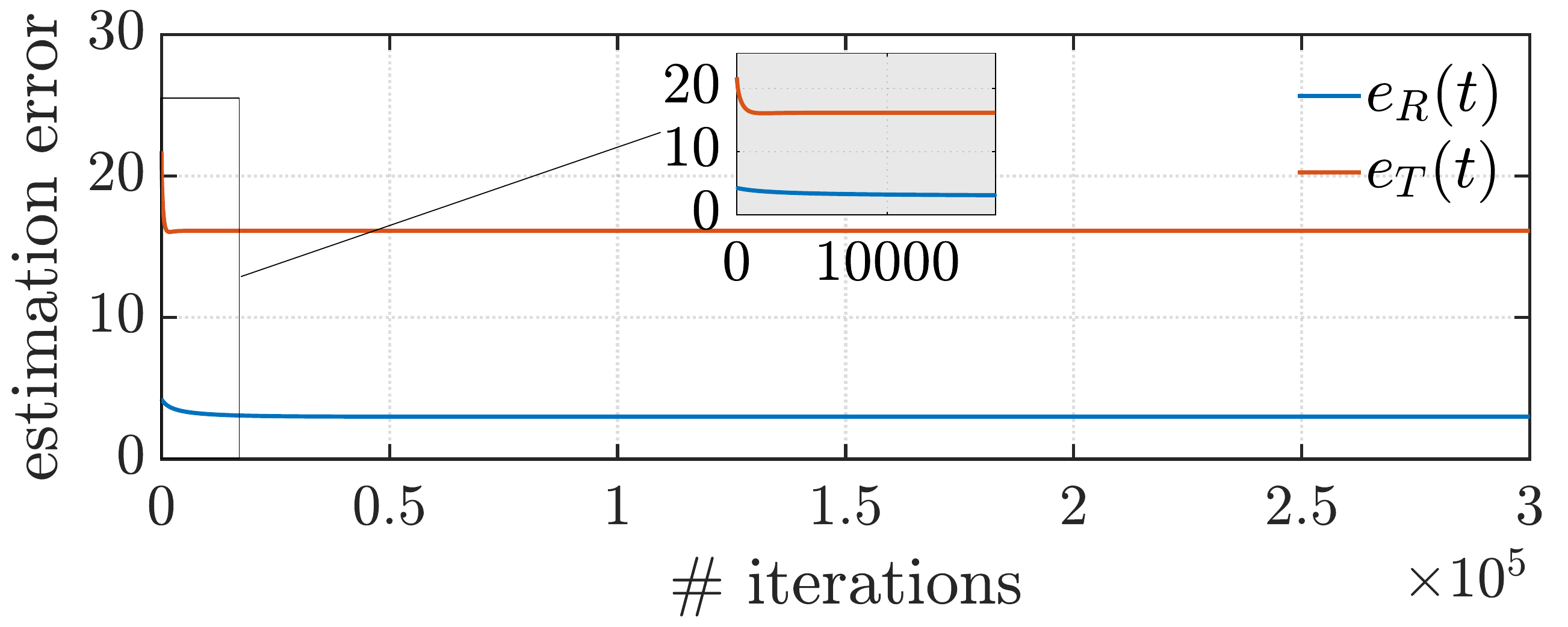}
\label{fig:TVBadCI_c}}
\end{center}
\caption{Performance of TV algorithm with bad initial conditions: (a) real cameras poses (blue), initial estimates (red) and final estimates (green); (b) trend of the cost function and its components in logarithmic scale; (c) trend of the average estimation error on orientations (blue line) and positions (red line).}
\vspace{-0.5cm}
\label{fig:TVBadCI}
\end{figure}

\begin{figure}[!t]
\begin{center}
\subfigure[]{ \includegraphics[width=0.36\textwidth]{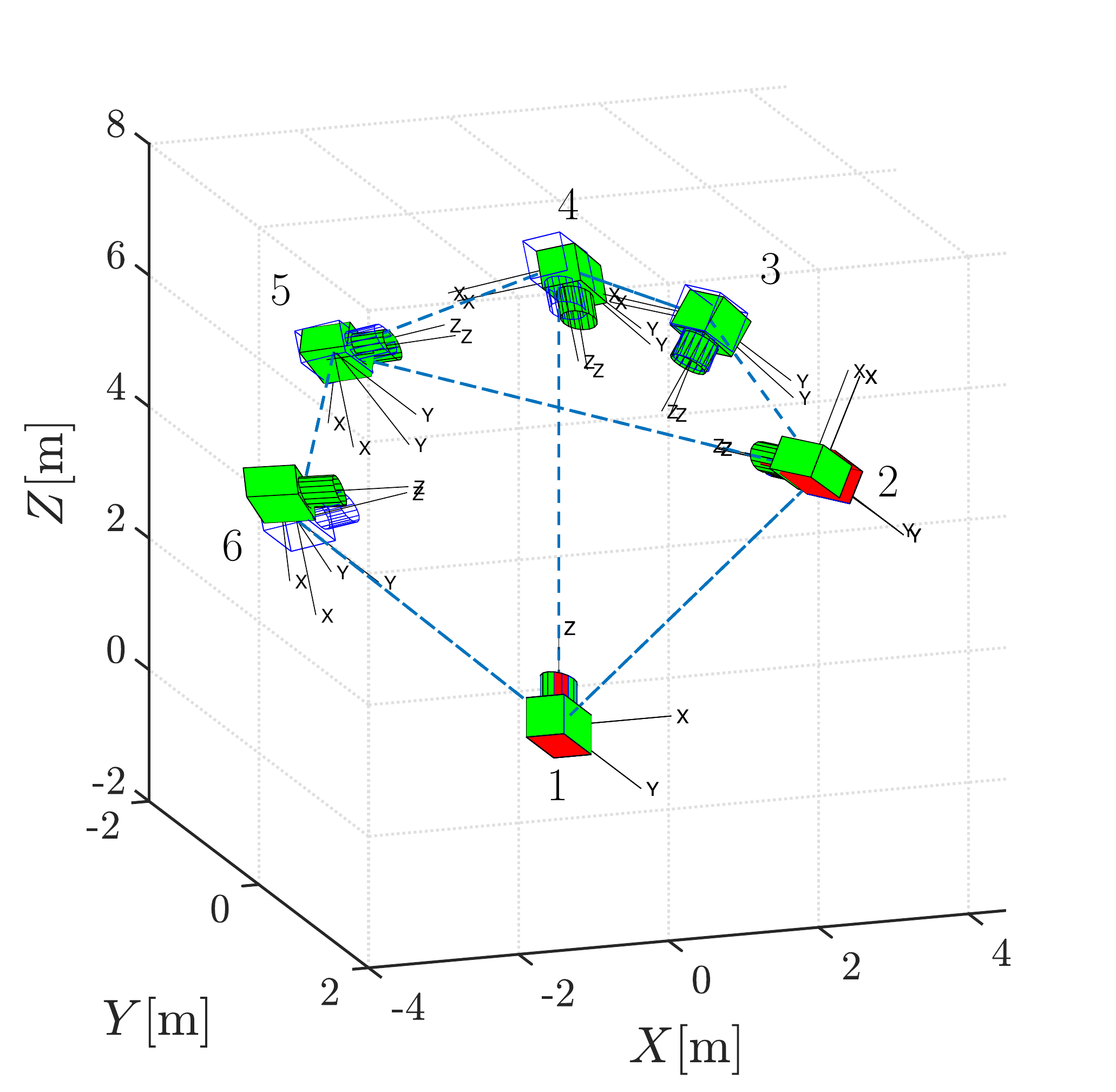} \label{fig:DQBadCI_a}}\vspace{-0.36cm}
\subfigure[]{
\includegraphics[scale=0.3]{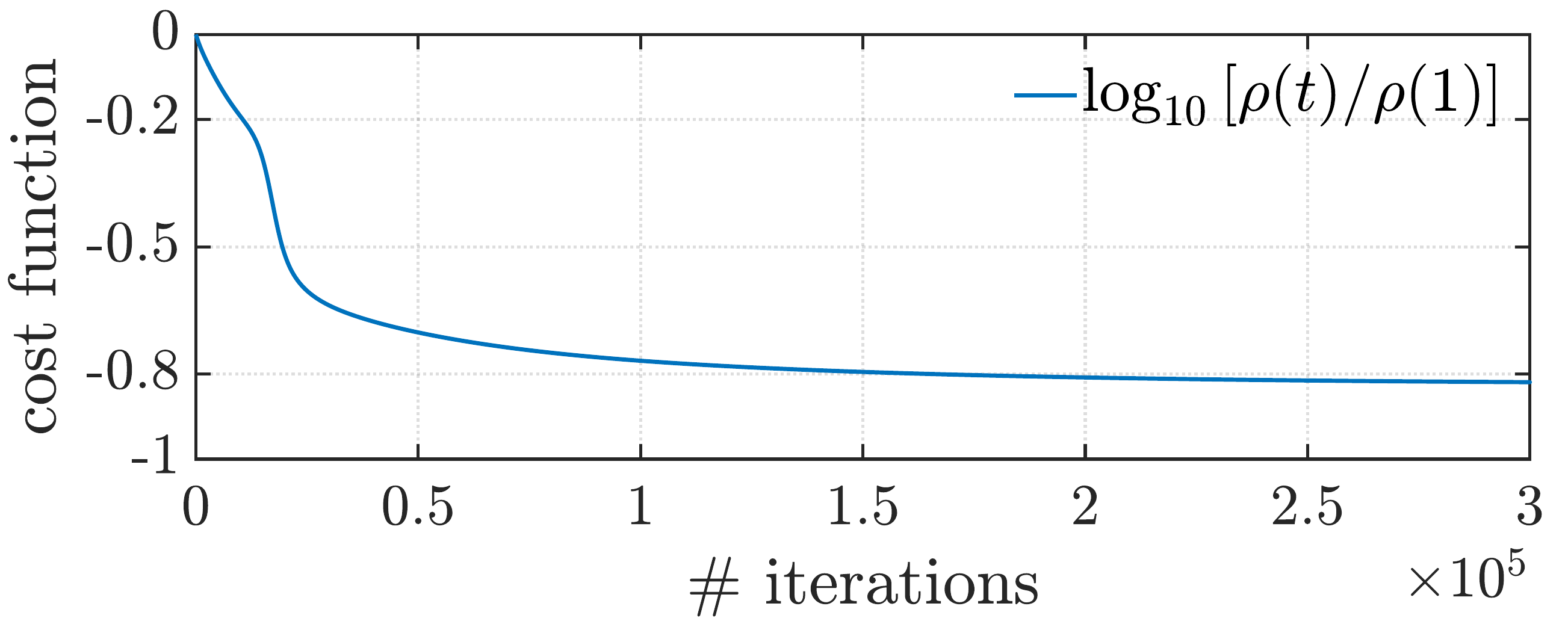}
\label{fig:DQBadCI_b}}\vspace{-0.36cm}
\subfigure[]{ \includegraphics[scale=0.3]{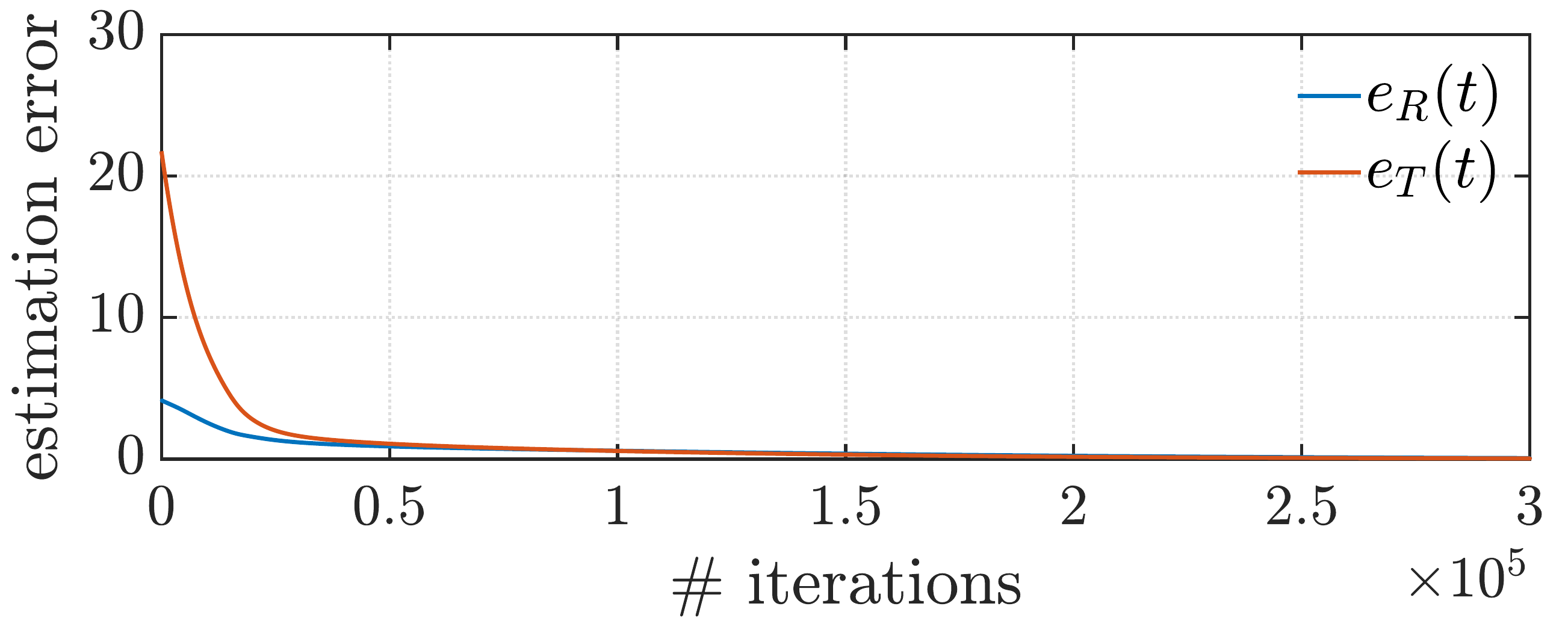}\label{fig:DQBadCI_c}}
\end{center}
\caption{Performance of DDQL algorithm with bad initial conditions:  (a) real cameras poses (blue), initial estimates (red) and final estimates (green), (b) cost function trend in logarithmic scale; (c) trend of the average estimation error on orientations (blue line) and positions (red line).}
\vspace{-0.5cm}
\label{fig:DQBadCI}
\end{figure}

%
%

\begin{figure*}[t]
\centering
\subfigure[low variance]{\includegraphics[scale=0.32]{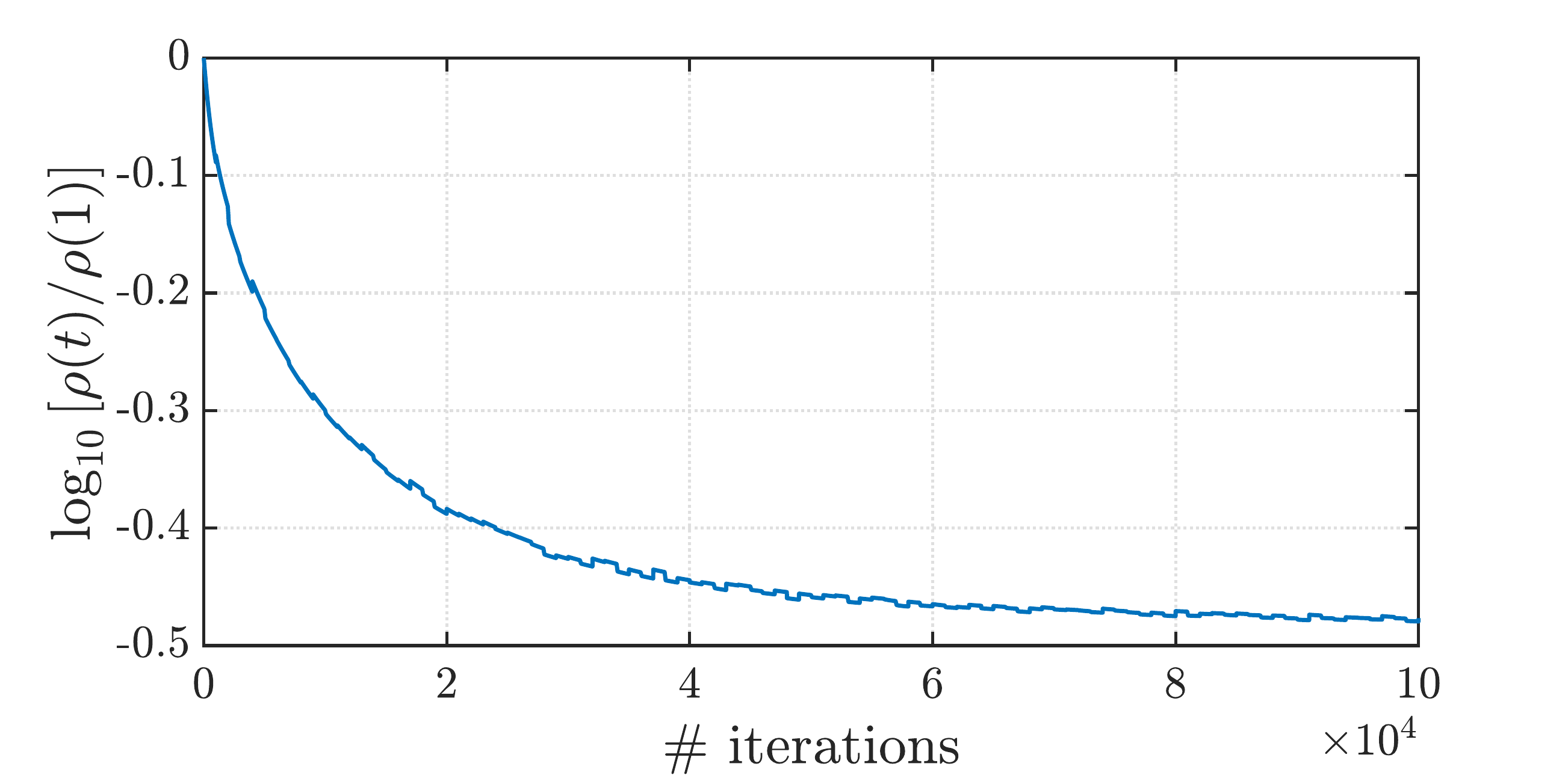} \label{fig:costFunctionDynGoodMeas}}
\subfigure[high variance]{\includegraphics[scale=0.32]{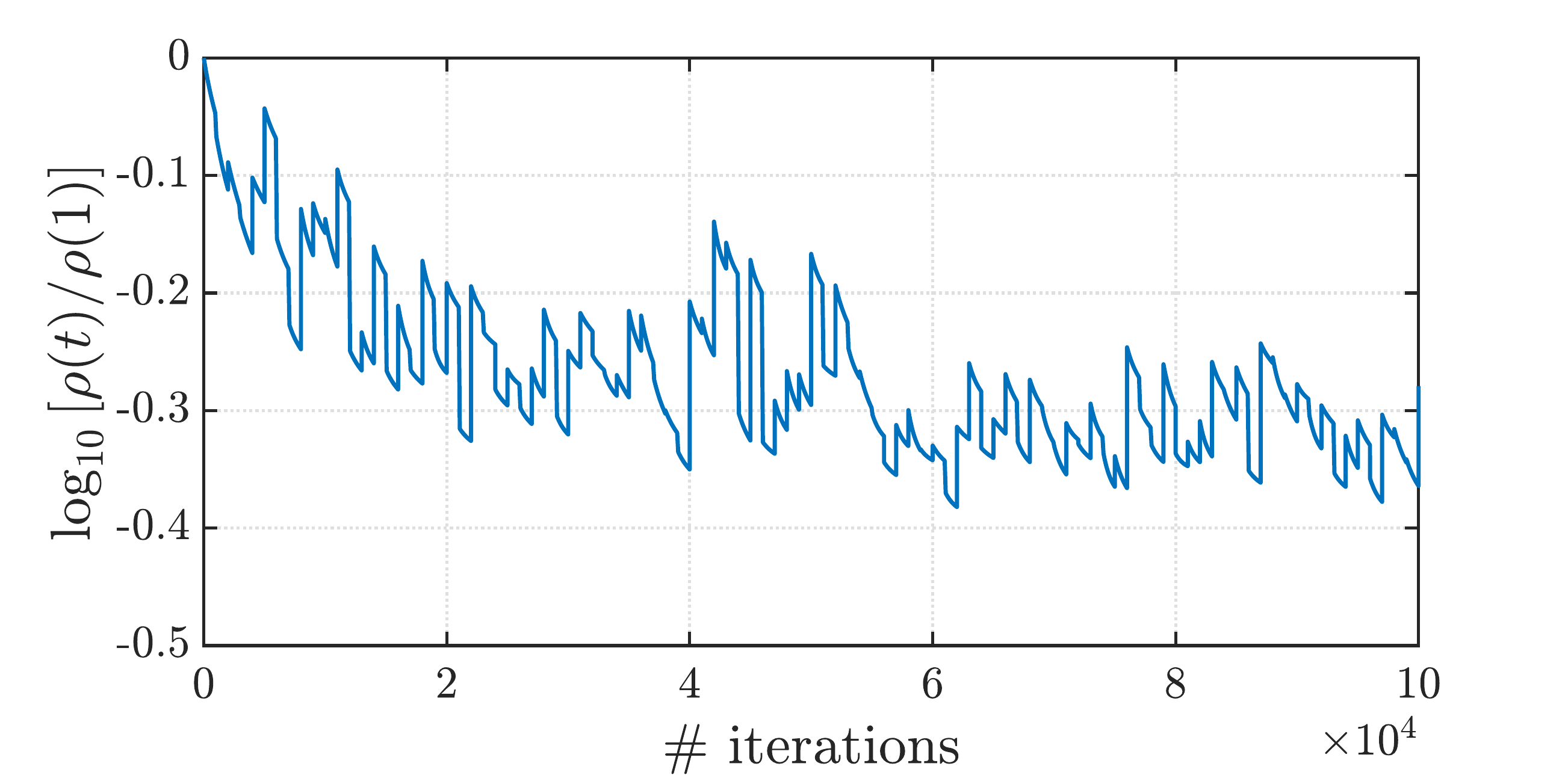}
\label{fig:costFunctionDynBadMeas}}
\caption{Cost function trend - vertical lines correspond to the beginning of a new localization step.}
\end{figure*}

The output of the proposed method, i.e., the final pose estimates reported in green in Fig.~\ref{fig:standardCase_a}, approaches the ground truth, although a small error affects the cameras orientation estimate along the $Y$-axis. This observation can be justified by the fact that the noise on the relative rotations is assumed to have a larger variance along this axis, nonetheless the estimation error results to be well distributed between the orientations and positions estimates. It is  also worth to notice that the cost function~\eqref{eq:cost_fun}, whose trend is depicted in Fig.~\ref{fig:standardCase_b}, decreases by almost $16 \%$ from its initial value: this represents a sensible minimization.

\subsection{Comparison between TV and DDQL algorithm}


To highlight the advantages of adopting $\mathbb{DH}_u$ instead of $\mathbb{SE}(3)$ as pose manifold, we compare the performance of the DDQL algorithm w.r.t. the TV one. In particular, we consider the previously described network but we account for a noise-free setup wherein the relative measurements correspond to the real relative poses. Moreover, to stress the robustness of Alg.~\ref{algo:DQreformulation} w.r.t. the initial conditions when compared to the approach described in~\cite{TronVidal}, we suppose to initialize all the poses estimates so that they correspond to the true pose of one of the devices (except for the first camera that is initialized to its own true pose).


Fig.~\ref{fig:TVBadCI} shows the performance of the TV-algorithm. We observe that the final estimates (green cameras in Fig~\ref{fig:TVBadCI_a}) are very far from the real poses (blue profiled cameras in the same figure). This behaviour is consistent with the fact that the cost function~\eqref{eq:cost_fun} and its components ($\rho_R(\cdot)$ and $\rho_T(\cdot)$ introduced in~\eqref{eq:cost_TV}) attain a local minimum, as depicted in Fig~\ref{fig:TVBadCI_b}. Fig.~\ref{fig:TVBadCI_c} describes the behaviour of $e_R(\cdot)$ and $e_T(\cdot)$, namely the mean estimation errors on orientations and positions, respectively. These quantities are computed according to the following expressions
\begin{align}
& e_R(t) = \dfrac{1}{n} \sum_{i\in \mathcal{V}} \lVert \mathbf{R}_i - \hat{\mathbf{R}}_i(t) \rVert_F^2, \\
& e_T(t) = \dfrac{1}{n} \sum_{i\in \mathcal{V}} \lVert \mathbf{p}_i - \hat{\mathbf{p}}_i(t) \rVert^2,
\end{align}
where $\hat{\mathbf{R}}_i(t) \in \mathbb{SO}(3)$ and $\hat{\mathbf{p}}_i(t) \in \mathbb{R}^{3}$ denote the $i$-th camera estimation of its real rotation matrix ${\mathbf{R}}_i \in \mathbb{SO}(3)$ and~position vector ${\mathbf{p}}_i \in \mathbb{R}^{3}$ at iteration $t$, respectively, and $\lVert \cdot \rVert_F^2$ indicates the Frobenius norm. As expected, both $e_R(\cdot)$ and $e_T(\cdot)$ decrease only during the first iterations, namely before the local minimum is reached, moreover, in both cases the reduction w.r.t. initial value is moderate. 
Fig.~\ref{fig:TVBadCI_b} and Fig.~\ref{fig:TVBadCI_c}  highlight one of the main drawbacks of TV-algorithm: position estimates are influenced by rotation estimates, hence, $\rho_T(\cdot)$ and $e_T(\cdot)$ experience a smaller decrease compared to $\rho_R(\cdot)$ and $e_R(\cdot)$. Remarkably, this result is independent on the specific initialization, being related to the adopted sequential optimization strategy.

Fig.~\ref{fig:DQBadCI} reports the results of the DDQL algorithm. It clearly appears that this approach is more robust w.r.t. the initial conditions as compared to the TV algorithm: the final poses estimates are very close to the real ones (Fig.~\ref{fig:DQBadCI_a}) and the cost function significantly decreases (Fig.~\ref{fig:DQBadCI_b}). In addition, Fig.~\ref{fig:DQBadCI_c} shows that both orientation and position errors\footnote{In this case  $\hat{\mathbf{R}}_i(t)$ and  $\hat{\mathbf{p}}_i(t)$ are extracted from  $\hat{\mathbf{d}}_i(t)$. Indeed, given $\mat{d} = {\mathbf{q}}_r+ \epsilon {\mathbf{q}}_d$, the real quaternion is associated to the rotation matrix
$
 \mathbf{R}(\mathbf{q}_r) = \mathbf{I}_3 + 2 q_{r,0} [\bar{\mathbf{q}}_r]_{\times} + 2[\bar{\mathbf{q}}_r]_{\times}^2
$, according to the Rodrigues formula. Furthermore, from the quaternion composition $2\mathbf{q}_d \circ \mathbf{q}_r$, the  position vector can be extract employing~\eqref{eq:RTtoDH}.} significantly decrease w.r.t. their initial values. In particular, it can be observed that both $e_R(\cdot)$ and $e_T(\cdot)$ tend to zero. As a consequence, at the steady-state we have that $e_R(\cdot) \approx e_T(\cdot)$: this is due to the balanced estimation error distribution, i.e., there is not a larger error accumulation in position estimates than in orientation ones, contrarily to the TV case. 

\subsection{Multi-sampling Measurements Case}

To establish the goodness of the DDQL algorithm in case of time-varying measurements, we consider the camera network previously described under the hypothesis that new relative poses measurements are collected after $Ts = 10^3$ iterations requiring the implementation of Alg.~\ref{algo:multiMeas}.  
Since the performance of the procedure depends on the variance of the measurements noise, we firstly assume a standard deviation of $\sqrt{5}$deg for rotations along $X$ and $Z$ axes, $5$deg for rotations along $Y$-axis and $\sqrt{0.005}$m for each position component (low noise variance case). 
Then we increase the standard deviation on rotations along $X$ and $Z$ axes to $10$deg, $100$deg for rotations along $Y$-axis and $\sqrt{0.5}$m for each position component (high noise variance case).

The trend of the cost function is reported in Fig.~\ref{fig:costFunctionDynGoodMeas} for the first case and in Fig.~\ref{fig:costFunctionDynBadMeas} for the second one. We note that, as expected, the low noise variance implies a quasi-smooth behavior of $\rho(\cdot)$, while, when the variance increases, the cost function decreases w.r.t. the initial quantity but wide oscillations make its trend irregular and slow.

\section{Conclusions}\label{sec: conclusions}

In this work, we propose a distributed algorithm solving the self-localization problem for a VSN. The original and innovative aspect of the provided method, termed DDQL algorithm, relies on the exploitation of the (unit) dual quaternion algebra to represent the pose of the devices in the network, which yields a modification to the optimization framework w.r.t the current literature. 
In particular, as compared to the approach described in~\cite{TronVidal}, that constitutes the starting point for the designed solution, the DDQL method does not  require to split the pose estimation problem into two  (rotation and position) contributes entailing a better distribution of the estimation error over the two pose components. Numerical simulations support this observation and, in addition, prove that DDQL inherently exhibits an increased robustness w.r.t. the estimation initial conditions.





\bibliographystyle{IEEEtran}
\bibliography{IEEEfull,biblio}

\end{document}